\documentclass[fleqn,usenatbib]{mnras}

\usepackage{newtxtext,newtxmath}

\usepackage[T1]{fontenc}

\DeclareRobustCommand{\VAN}[3]{#2}
\let\VANthebibliography\thebibliography
\def\thebibliography{\DeclareRobustCommand{\VAN}[3]{##3}\VANthebibliography}

\usepackage{graphicx}	
\usepackage{amsmath}	
\usepackage{subcaption}

\newcommand{\lw}{\textit{Lightweaver}}

\title[\ion{C}{i} Rydberg line diagnostics]{Diagnostics for the solar chromosphere using neutral carbon Rydberg lines}

\author[R.P. Dufresne et al.]{
R.P. Dufresne,$^{1}$\thanks{E-mail: rpd21@cam.ac.uk}
C.M.J. Osborne$^{2}$
and G. Del Zanna$^{1,3}$
\\
$^{1}$DAMTP, University of Cambridge, Wilberforce Road, Cambridge CB3 0WA, UK\\
$^{2}$SUPA School of Physics and Astronomy, University of Glasgow, Glasgow G12 8QQ, UK\\
$^{3}$School of Physics and Astronomy, University of Leicester, Leicester LE1 7RH, UK\\
}

\date{Accepted XXX. Received YYY; in original form ZZZ}

\pubyear{\the\year{}}

\begin{document}
\label{firstpage}
\pagerange{\pageref{firstpage}--\pageref{lastpage}}
\maketitle

\begin{abstract}
Diagnostics for the solar chromosphere are relatively few compared to other parts of the atmosphere. Despite this, hundreds of Rydberg lines emitted by neutrals in this region have been observed at UV wavelengths. Here, we investigate their diagnostic potential by modelling the lines emitted by neutral carbon using recent atomic data. We use the radiative transfer code \lw\ to explore how they form and how they respond to temperature, density and micro-turbulent velocity perturbations in the atmosphere. To simplify the modelling, we investigate lines emitted from levels with principal quantum number $n\geq10$, which are expected to be in Saha-Boltzmann equilibrium with the ground state of the singly-charged ion. Optical depth effects are apparent in the lines and their response to atmospheric perturbations suggest that they will be useful in reconstructions of the atmosphere using inversions. The study opens the way for using many such lines emitted by multiple elements over a range of heights, a large number of which will be observed by the forthcoming Solar-C EUV High-throughput Spectroscopic Telescope (EUVST).

\end{abstract}




\section{Introduction}

Diagnostics for the solar chromosphere have generally been focussed around a certain set of lines, including \ion{Mg}{ii} h and k \citep[such as][]{uitenbroek1997}, \ion{Ca}{ii} H and K \citep[see][for an early review]{linsky1970}, H$\alpha$ \citep[such as for flares by][]{heinzel1987}, and Ly-$\alpha$ \citep{basri1979} for the upper chromosphere; and, \ion{Mg}{ii} k$_1$, \ion{Ca}{ii} K$_1$ \citep{vernazza1981}, \ion{Mg}{i} b \citep{quintero2018}, \ion{Na}{i} D and the \ion{Ca}{ii} infra-red triplet \citep{rutten2011}, along with some continua and molecular lines, for the lower chromosphere. By contrast, there are numerous absorption lines in the photosphere below and emission lines from the transition region and corona above. Any additional diagnostics which could elucidate conditions in the complex conditions of the chromosphere would be a welcome tool for observers and modellers alike.

Lines which are formed by collisional excitation are unlikely to exist in this temperature range. The most suitable alternative would be lines which form through recombination from a singly-charged ion following photo-ionization of the neutral. Modelling lines from such highly excited levels, however, would involve complex atomic models for a correct non-LTE treatment, that is, for departures from local thermodynamic equilibrium (LTE). For example, \citet{lin2017} modelled the \ion{C}{i} 1355.8\,\AA\ observed by IRIS \citep{depontieu2014}, which is emitted from the highly-excited $2s^2\,2p\,4f\;^1D$ state. They started with a model atom containing 168 levels. They reduced this to an atom with 26 levels by only including the singlet states explicitly; all transitions involving triplets were totalled and included via the ground state. Such models are time-consuming to build, even if sufficient data is available. 

By contrast, it is well-known that for levels in Saha-Boltzmann (SB) equilibrium the atomic modelling is vastly simpler and such difficulties are avoided. The H model by \citet{storey1995} and the He model by \citet{delzanna2020he} indicate that levels with principal quantum number of approximately $n=10$ and higher should be in SB equilibrium with the ground state of the singly-charged ion in conditions present in the quiet solar chromosphere. At densities relevant for active regions and flares energy levels with even lower $n$ may be in SB equilibrium.


The observations of \citet{sandlin1986} and \citet{parenti2005}, for example, show that there are hundreds of lines emitted from Rydberg states at far-UV wavelengths in the solar atmosphere. These start with \ion{N}{i} at 853.1\,\AA, continue with \ion{O}{i}, \ion{C}{i} and \ion{S}{i} and finish with \ion{Si}{i} at wavelengths beyond 1521.0\,\AA. (Rydberg lines from \ion{Mg}{ii}, \ion{Ca}{ii} and \ion{Fe}{ii} are also observed at these wavelengths, but it is not certain from which $n$ their Rydberg levels would be in SB equilibrium.) \citet{dufresne2025lwatoms} found that the Si continua start forming below the temperature minimum, while the C continua form at the temperature minimum and upwards. Therefore, given the wide variety of elements that emit Rydberg lines, it can be expected that the lines will be emitted over a range of heights and could provide diagnostics where there are currently gaps. Modelling such levels in SB equilibrium would only require the relative populations of long-lived levels in the neutral and singly-charged species, the energies of the Rydberg states and the spontaneous decay rates for the relevant lines.

\begin{figure*}
\centering
\includegraphics[width=\textwidth]{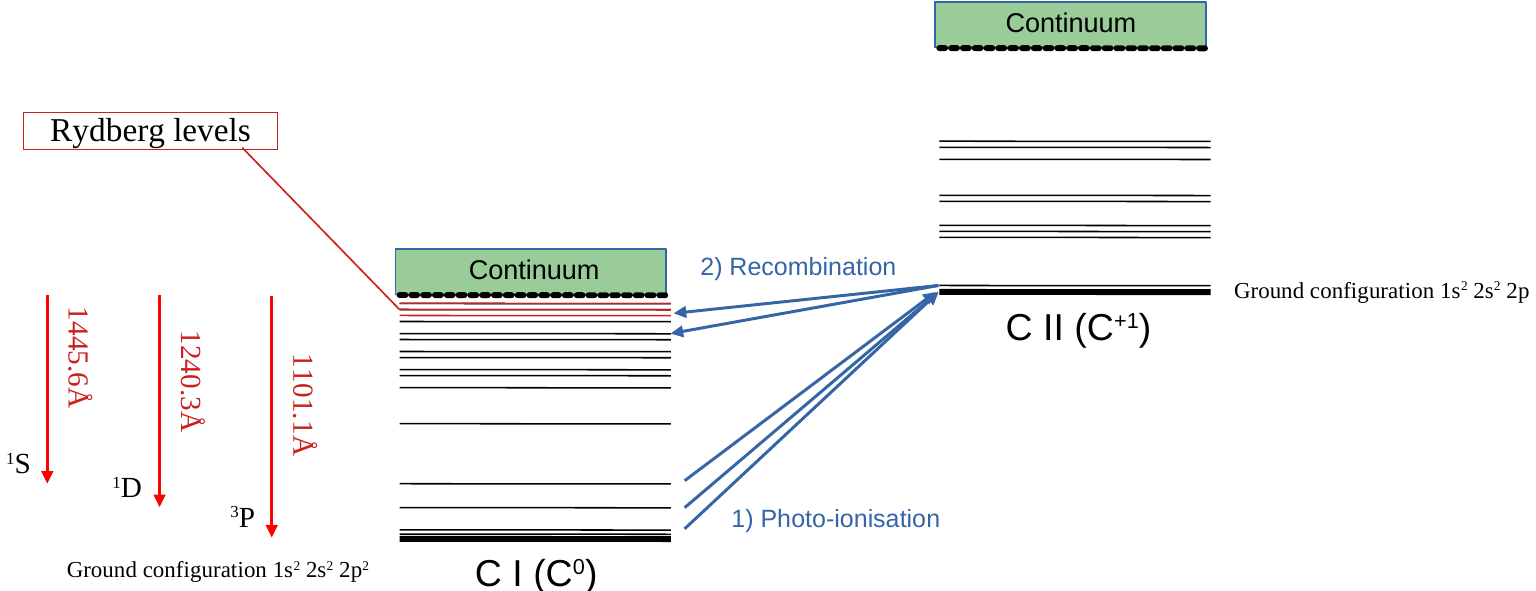}
\caption{Schematic illustration of the formation mechanism of \ion{C}{i} Rydberg lines, including the series limits at UV wavelengths.}
\label{fig:lineformation}
\end{figure*}

These diagnostics would also be useful in another respect. Inversions of chromospheric observations reconstruct key parameters like temperature stratification, electron density, line-of-sight velocities and magnetic field. One example is \citet{morosin2022}, who use \ion{H}{i}, \ion{Mg}{ii} and \ion{Ca}{ii} lines to reconstruct a plage chromosphere. This first requires understanding the response of the level populations to changes in the physical parameters and then, from this, understanding the response of the lines to the same perturbations. The speed of the inversions will depend on the ease with which the forward modelling can be carried out. Clearly, if the levels are in SB equilibrium their response will be straightforward because of their analytical dependence on temperature and density. All that remains is to know the relative ion fractions. Although the ion fractions form outside of SB equilibrium, they should change more slowly than level populations under small perturbations.

The aim of this paper is to study the formation of Rydberg lines and their response to changes in the atmospheric parameters. This will pave the way for their use as spectroscopic diagnostics and in inversions. All the required data for the Rydberg states of neutral carbon have been made available recently: both the atomic data for the lines themselves \citep{storey2023c1} and the improved atomic rates used in radiative transfer calculations to estimate ion fractions \citep{dufresne2025lwatoms}. The latter work did, however, highlight that atmospheric models still require improvement. \citet{storey2023c1} found evidence of opacity in the Rydberg lines. They carried out a simple escape fraction estimate for the lines and suggested that lines decaying to different lower levels, with series limits at 1101.1\,\AA, 1240.3\,\AA\ and 1445.7\,\AA, form at different heights in the lower chromosphere. Consequently, we use the \lw\ radiative transfer code here for this test case. Once their formation are understood, it opens up the way for using potentially hundreds of such lines in the UV range.

This study is particularly timely with the forthcoming EUV High-throughput Spectroscopic Telescope (EUVST) onboard Solar-C, which will provide high spatial and spectral resolution observations from the solar chromosphere to corona. It will be an ideal instrument with which to resolve the relatively narrow lines emitted by neutrals. The present analysis will also lay the foundation for those who wish to extract velocity field information from the complex line profiles observed during flares; an example being \citet{monson2024} who analysed this using \ion{Fe}{i} lines. More understanding of Rydberg line emission will help assess whether they have any impact on the total radiative losses from this region, which are important for magneto-hydrodynamic simulations and understanding heating mechanisms in the solar atmosphere \citep[see][for example]{dasilva2024}. These diagnostics and their impact on understanding this region will inform current and future large-scale models of the chromosphere, such as MURaM \citep{przybylski2022} and the Solar Atomspheric Modelling Suite (SAMS\footnote{https://sams-project-uk.github.io/}). Rydberg lines have also been observed in stellar spectra \citep[see][for further details]{hall2008lrsp, linsky2019}.

The next section of this article covers the methods used for the radiative transfer calculation, the atomic data used for the lines, how the line response functions are calculated and the observations used to compare the outputs. Section\;\ref{sec:results} present the synthetic spectrum compared to solar observations, analyses the contribution functions and shows the response functions. Finally, Sect.\;\ref{sec:concl} presents some conclusions.

\section{Methods}
\label{sec:methods}

Within the solar chromosphere, the populations of the states in neutral  carbon are driven primarily by photo-ionisation from the long-lived levels in C$^0$, followed by radiative and dielectronic recombination from the ground state of C$^+$. (A schematic illustration of the lines and the wavelength limits of each series of lines is shown in Fig.\;\ref{fig:lineformation}.) 
This is a non-linear problem, generally requiring a large-scale
collisional-radiative model with C$^0$ and C$^+$ including all atomic processes connecting states \citep[see, e.g. the helium model by][]{delzanna2020he} and non-LTE radiative transfer to solve the level populations and, hence, the ion balance \citep{rathore2015}. As recombination goes through high-lying Rydberg states and the number of energy levels scales with $n^2$, such atomic models can be very large. 
The radiative transfer calculation then becomes computationally demanding, in addition to the complexity in building such an atomic model in the first place. However, if one is interested in the populations of the Rydberg levels, this can be avoided. 
At the high densities of the solar chromosphere, collisional 
processes among the states and the C$^+$ ground state are so strong that 
the populations of the Rydbergs levels can be described by the Saha-Boltzmann equation and will be proportional to the number density of C$^+$. In other words, the Rydberg levels are in SB equilibrium with the ground state of C$^+$ and the calculations can be simplified considerably. 
%
%
Because the Rydberg levels are connected to C$^+$ they are not in SB equilibrium with the lower states of C$^0$, which means the Rydberg lines are formed in non-LTE conditions. 
The populations of Rydberg levels are relatively small and do not affect the population of the long-lived levels. Therefore, the radiative transfer calculation can be split up into two parts. Firstly, the calculation to solve the long-lived levels are carried out with a limited atomic model. In the second run, just the ground and metastable levels of C$^0$ and C$^+$ are included and their populations fixed at the values of the first run. The Rydberg levels are then added to the model and their populations fixed at Saha-Boltzmann values for the remainder of the calculation.

We use the \lw\ radiative transfer code \citep[][v0.13.0]{osborne2021lw} and the same setup as \citet{dufresne2025lwatoms}, namely steady-state, 1D radiative transfer with H, He, C, O, Mg, Al, Si, S and Fe in non-LTE and N, Ca and Ni in LTE. The new atomic models from that work for C, Si and S are used here. The emphasis there was on the continuum, and so the \citet{fontenla2014} cell model was used as the model atmosphere input in that work. In this case, lines are being modelled and so we use the network model from \citet{fontenla2014} instead; inter-granular network is expected to make the greatest contribution to line intensity \citep{mariska1992}. Partial redistribution (PRD) is used throughout both calculations for hydrogen Lyman-$\alpha$ and -$\beta$ lines. This is important for the Rydberg series at 1240\,\AA\ because it forms on the wings of Lyman-$\alpha$. 

For the Rydberg levels, we add all the data from \citet{storey2023c1} to the carbon model: level energies relative to the continuum and radiative decay rates. 
The populations for the Rydberg levels are calculated using the parameters from the model atmosphere and the Saha-Boltzmann equation:

\begin{equation}
    \frac{N_u}{N_p} \;=\; \frac{g_u}{2g_p} \sqrt{\left( \frac{h^2}{2\pi m_e kT_e} \right)^3} \;{\rm exp}\left( \frac{I_{up}}{kT_e}\right) \; N_e ~,
    \label{eqn:saha}
\end{equation}

\noindent where $N_u$ is the number density of a Rydberg level $u$ relative to the number density, $N_p$, of its parent $p$, $g$ is the statistical weight of a level, $T_e$ is the electron temperature, which is assumed to be the same as the ion temperature, $I_{up}$ is the ionisation potential of the Rydberg level relative to its parent, and $N_e$ the electron number density.

\begin{figure*}
\centering
\includegraphics[width=\textwidth]{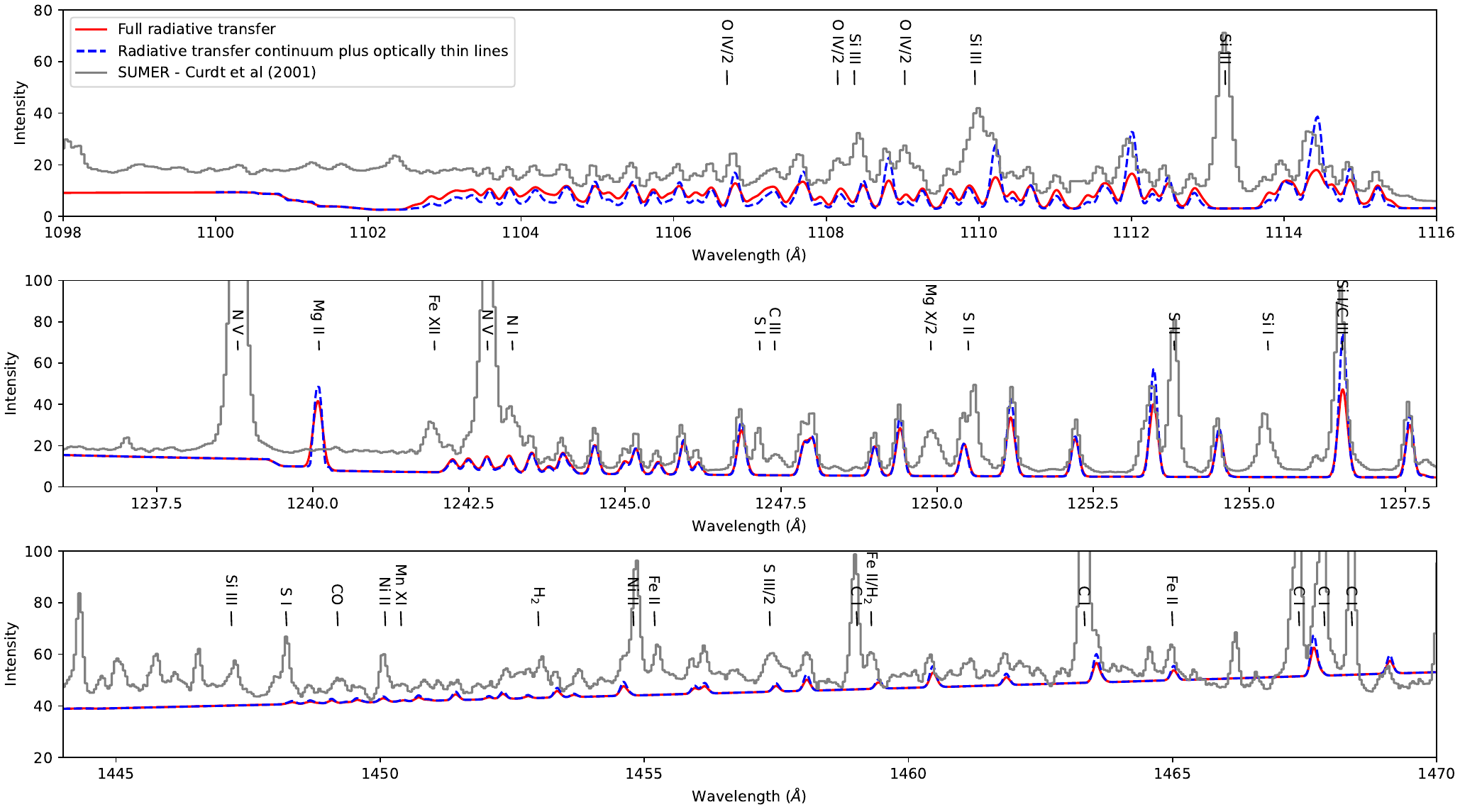}
\caption{Comparison of \ion{C}{i} Rydberg line intensities from the radiative transfer calculation with observations and with the relative intensities from the optically thin calculations of \citet{storey2023c1}. The other main lines contributing to the spectrum that have been identified and are not being modelled here are indicated by black labels. The `/2' in the labels denotes a second order line in SUMER. All intensities are in erg\,s$^{-1}$\,cm$^{-2}$\,sr$^{-1}$\,\AA$^{-1}$.}
\label{fig:rtlines}
\end{figure*}

The emergent intensity of a line is given by

\begin{equation}
I_0 (\mu,\nu) = \int^{\infty}_0 \;
 S (\tau_\nu , \mu, \nu)\,e^{-\tau_\nu}\; d\tau_\nu /\mu,
\end{equation}

\noindent where $\mu=cos\,\theta$, $\theta$ is the viewing angle with respect to the surface normal, $S (\tau_\nu , \mu, \nu)$ is the source function. Hereafter, the source function will be abbreviated by $S_\nu$ and $S_\nu= \frac{\eta_\nu}{\chi_\nu}$, where $\eta_\nu$ is the emissivity, $\chi_\nu$ the opacity. $\tau_\nu$ is the monochromatic optical depth, defined by $d\tau_\nu=-\chi_\nu\,dz$, where z is the vertical height through the atmosphere. The integrand is the contribution function and will be abbreviated by $C_I$.


Here, we define the response function $R_q(\mu, \nu, z)$ to a functional perturbation $\delta q(z)$ of atmospheric parameter $q$ stratified along $z$ as the relative change in outgoing intensity due to this perturbation

\begin{equation}
\frac{\delta I_0(\mu, \nu)}{I_0(\mu, \nu)} = \int_{z_\mathrm{max}}^0 R_q(\mu, \nu, z) q(z) dz \; ,
\end{equation}

\noindent where $z_{\max}$ is the maximum depth of the model \citep{ruizcobo_sir:1992}. 



Since we are using a discrete atmosphere this is carried out by perturbing at one point, $k$, in the atmosphere the parameter $q$ by an amount $\pm \delta q_k$ and calculating the response function to this perturbation as

\begin{equation}
 R_{q_k} (\mu,\nu) = \frac{1}{I_0 (\mu,\nu)}\, \frac{I_0 (\mu,\nu,q_k+\delta q_k)\,-\,I_0 (\mu,\nu,q_k-\delta q_k)}{2\delta q_k} \; .
\end{equation}

\noindent For the response functions, we perturb the temperature and density by $\pm$2\,\% and the micro-turbulent velocity by $\pm100$\,m\,s$^{-1}$.

We follow the assessment of \citet{storey2023c1,dufresne2025lwatoms} and use the SOlar Heliospheric Observatory (SOHO) Solar Ultraviolet Measurements of Emitted Radiation (SUMER) instrument for the comparison of the synthetic spectrum with observations. \cite{curdt2001} provided a complete line list of wavelengths from 680 to 1611\,\AA, merging observations obtained over a time span of several hours on 1997-04-20. By examining other data from SUMER and the overlapping regions within these observations (of about 8\,\AA), very little variability between exposures of the lines from neutrals was present. Therefore, lines at very different wavelengths can be safely compared. Above 1475\,\AA\ observations were taken from the uncoated Detector A and a step is noted in the observed intensity. Also above that wavelength the second-order contribution from the H Lyman continuum becomes significant. Both of these issues are at wavelengths longer than our analysis here, and so we use the \citet{curdt2001} published data. The line profiles of the neutrals are mostly instrumental in SUMER, which means we broaden our synthetic line profiles with a FWHM of 0.11\,\AA, the value preferred by \citet{rao2022}, using the \textsc{PyAstronomy}\footnote{https://github.com/sczesla/PyAstronomy} package \citep{pya}.

\section{Results}
\label{sec:results}

\subsection{Line formation}
\label{sec:lines}

It is customary in the analysis of lines in the transition region and corona to consider them separately from the continuum. Because of limitations in modelling the C continuum intensities found in \citet{dufresne2025lwatoms}, this section discusses only the intensities of the lines themselves separately from the continuum. Issues with reproducing the C continuum and how that may affect the Rydberg line intensities are discussed later, in Sect.\;\ref{sec:cont}.

The synthetic spectrum from the radiative transfer calculation at the wavelengths of the carbon Rydberg lines is shown in Fig.\;\ref{fig:rtlines}. 
The lines are broadened with a Gaussian profile of 0.13\,\AA, which includes the thermal and instrumental broadening.
The intensities of the lines themselves, if one subtracted the continuum, are in very good agreement with observations for the two series at shorter wavelengths. The results also capture the line shapes and blends very well. It is only the series decaying to the $^1S$ term, at wavelengths longer than 1450\,\AA, that appear further from observations. In this longer wavelength region there are many molecular and unidentified lines, and so it is not straightforward to know in the observations which are the carbon lines and whether they are unblended. The most likely candidates in this region are the lines close to 1456.0\,\AA. If they are unblended, then the synthetic line intensities are approximately a factor of two to three below observations. More discussion will be given in Sect.\;\ref{sec:cont} concerning the extent to which an under-prediction in the C$^+$ fractions could be causing this.


We also show in Fig.\;\ref{fig:rtlines} the results from the optically thin calculations of \citet{storey2023c1} added to the synthetic continuum calculated by \lw\ in the present work. To provide a rough comparison of the new atomic data with solar observations, \citet{storey2023c1} introduced ad-hoc scaling factors to normalize the line emissivities, one for each series.
The emissivities were calculated at one point in the atmosphere because the relative intensities in optically thin conditions change little within each series no matter the height at which the lines form. In other words, \citet{storey2023c1} did not use an emission measure to integrate intensities along the line of sight for the optically thin calculation. 

It is clear that the RT \lw\ calculations represent a significant
improvement over the previous calculations: the lines and the continua are calculated self-consistently from the same atmospheric model,
and no scaling factors have been introduced to normalise the emissivities of the three series. 
The comparison also shows the improvement due to the RT calculations in the lower $n$ states (those at longer wavelengths within each series), where the optically thin results over-predict the observed intensities.

On a side note, \citet{storey2023c1} also included a calculation of emergent intensities for a limited set of lines around 1115\,\AA\ and 1256\,\AA\ using the escape probability formalism and the \citet{vernazza1981} model C atmosphere. This showed improved agreement in the relative intensities of the stronger lines compared to the optically thin results, but there were still disagreements of around factors of two for those lines.
 
The differences between the full RT calculation, the optically thin results and observations lie in the stronger lines in each series as $n$ decreases, that is, as wavelength increases. The populations of levels with the same parent and total angular momentum, $J$, are within 10\% of each other between $n=10$ and $n=20$. Also, if the levels at $n=10$ were not in SB equilibrium their populations would actually be below SB equilibrium values and their lines intensities would be lower than the predictions here. Therefore, it cannot be changes to level populations that are affecting the results. Another source of the discrepancy could be the radiative decay rates. This is harder to assess because there were no other, reliable atomic calculations for $n\geq10$ to compare the data. If there were discrepancies in the radiative decay rates for transitions from the $J=3$ states, say, it is likely that a systematic error would be seen for decay rates from all such states. The \citeauthor{storey2023c1} rates were well within 10\% of other calculations for $n<10$, and there does not seem to be an inconsistency between theory and observations for decays closer to the continuum edges.

\begin{figure*}
\centering
\includegraphics[width=\textwidth]{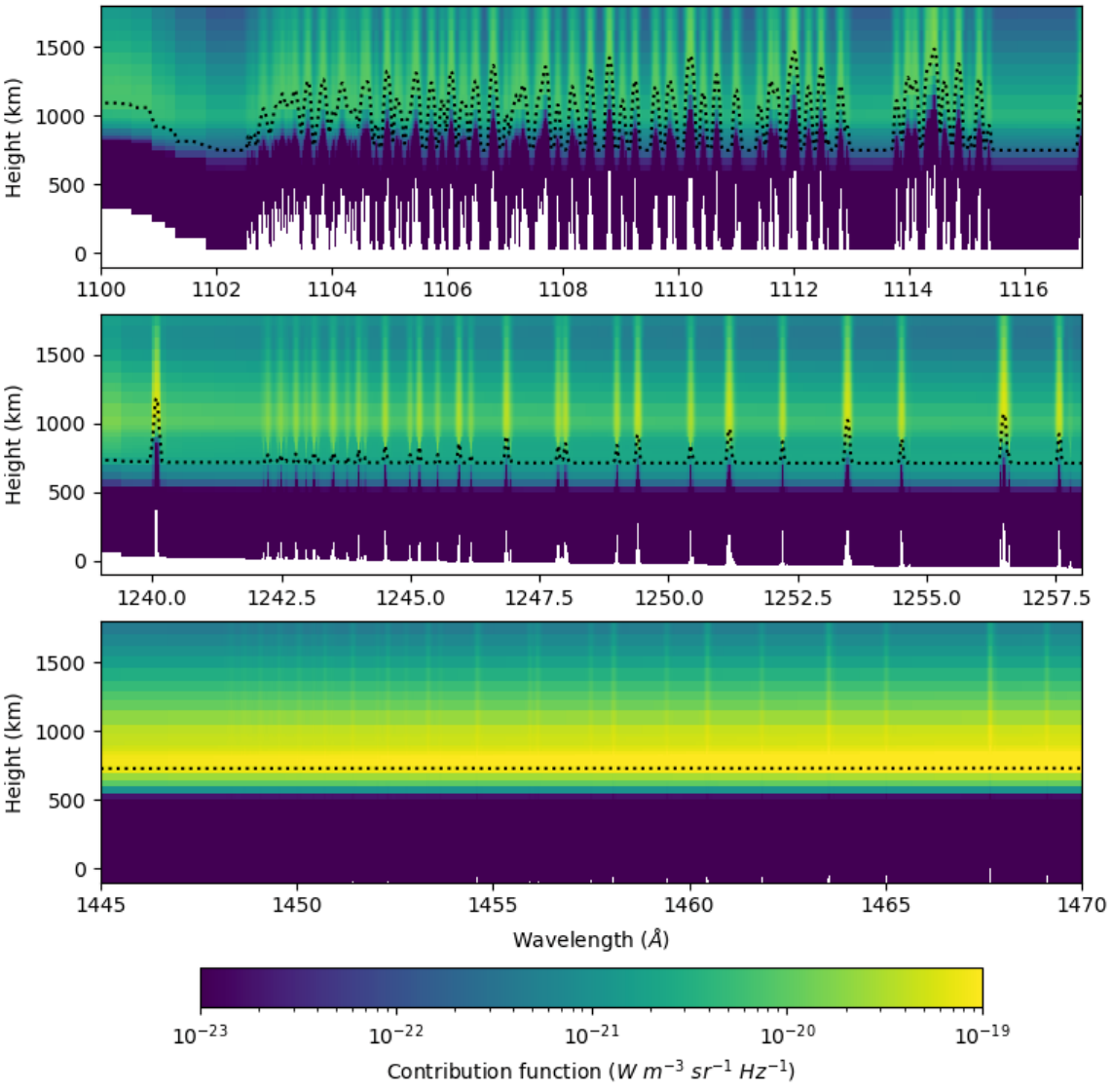}
\caption{Contribution functions for all three Rydberg series. The $\tau=1$ layer (black dotted line) is also shown.}
\label{fig:allcontribs}
\end{figure*}


\begin{figure*}
\centering
\begin{subfigure}{0.9\textwidth}
\includegraphics[width=\linewidth]{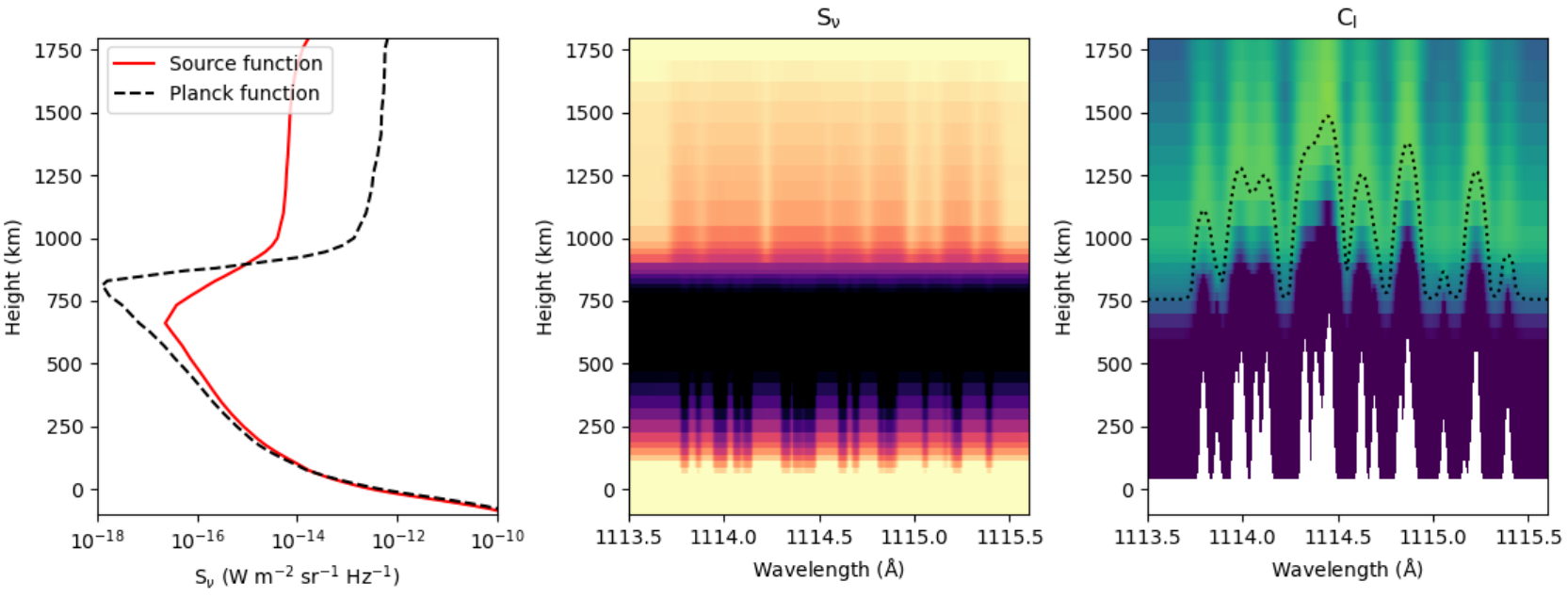}
\label{fig:1115sect}
\end{subfigure}
\hfill
\begin{subfigure}{0.9\textwidth}
\includegraphics[width=\linewidth]{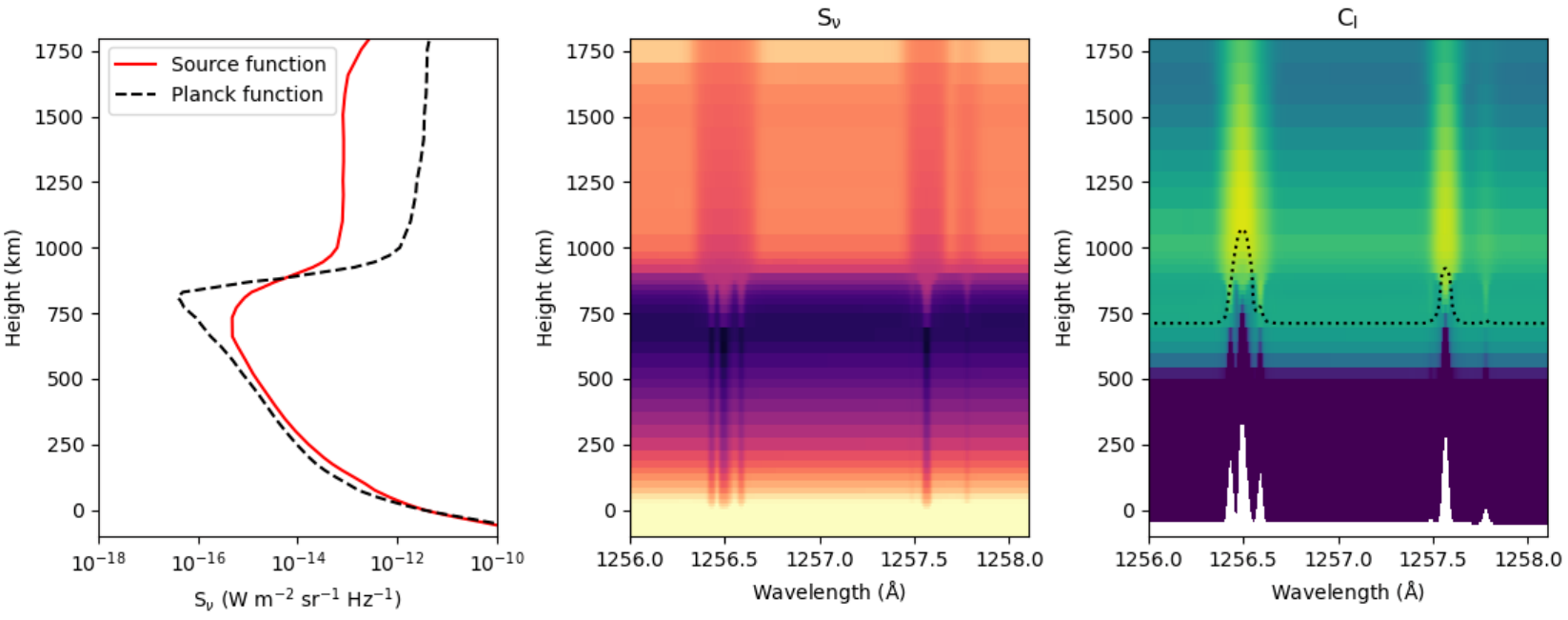}
\label{fig:1257sect}
\end{subfigure}

\caption{The left plots show the source and Planck functions with height at line centres 1114.46\,\AA\ (upper plots) and 1256.50\,\AA\ (lower plots), the middle plots show the source functions with respect to wavelength and height, and the right plots show the contribution functions. The $\tau=1$ layer is shown on the contribution function plots by black dotted lines. Brighter colours indicate greater values of the source and contribution functions. Both sets of lines are emitted from upper levels with $2s^2\,2p\,10d,11s$ configurations.}
\label{fig:source_fns}
\end{figure*}

Overall, then, these factors suggest that opacity is the main factor for the synthetic intensities being below observations for the stronger lines. A further clue to what may be going on comes from the fact that the series decaying to the $^1D$ and $^1S$ states are both below observations in the RT calculation. This is obviously because the C$^+$ populations are too low, which will be discussed further in Section\;\ref{sec:cont}. This causes the level populations and line intensities to be lower because the levels are directly correlated with the C$^+$ ground. A corollary is that opacity will be greater in the lines because the C$^0$ populations are higher. This will affect not only the stronger lines in each series, but will also affect all the lines decaying to the ground state more than the lines from the other series.

\subsection{Contribution functions}
\label{sec:contrfn}

In Fig.\;\ref{fig:allcontribs} we show the contribution functions for all lines emitted from levels in SB equilibrium in all three series, for the purpose of making more general comments about line formation. We will give more details about individual lines later. Starting with the series at 1445-1470\,\AA, it is clear that the greatest contribution to emission from both the lines and continua is at about 750\,km (3865\,K). This is slightly lower in height than the temperature minimum in the \citet{fontenla2014} atmosphere at 800\.km (3750\,K). Contributions to the emergent intensity begin right at the $\tau=1$ layer and the lines continue emitting at heights above 1800\,km. The series at 1240-1260\,\AA\ contribute most to the emergent intensity at heights of around 1000\,km (5890\,K). The $\tau=1$ layer is lower in the atmosphere for lines emitted from higher $n$, that is, at shorter wavelengths and closer to the series limit, than for the stronger lines in the same series. Consequently, lines within this series are emitted at different heights and, overall, are emitted from higher in the atmosphere than the lines at 1445-1470\,\AA. In the case of the lines observed at 1100-1116\,\AA, because there are three lower states to decay to ($2s^2\,2p^2\;^3P_{0,1,2}$), there is a much wider variation in intensity in the lines and heights over which they form. The strongest lines at lower $n$, say at around 1115\,\AA, have a $\tau=1$ layer at about 1500\,km (6300\,K), appreciably higher than the continuum, which emits at similar temperatures to the continuum at 1245\,\AA.

We focus now, and for the response functions in Sect.\;\ref{sec:response_fns}, on a few lines so that the effects can be seen in more detail. We choose the lines emitted from the $2s^2\,2p^2\,10d,11s$ upper levels. The individual lines with their wavelengths, upper and lower levels, and emissivities from \citet{storey2023c1} for all three series are given for reference in Tab.\;\ref{tab:emiss}. These particular levels are chosen because they emit the lines observed by the forthcoming EUVST instrument and they were also the focus of the earlier work. Figure\;\ref{fig:source_fns} shows the source, Planck and contribution functions for the lines decaying to the $^3P$ and $^1D$ lower levels. In both cases, it is seen that the source function follows that of the Planck from the photosphere until just below the temperature minimum, which is located at 800\,km. Chromospheric and transition region radiation is able to penetrate to the temperature minimum region, causing an over-ionisation of C$^0$ and a higher source function. However, the emission does not escape because of line and continuum opacity. At higher heights the source function drops below that of the Planck function.

It is clear from the brighter colours in the middle plot of Fig.\;\ref{fig:source_fns} that the source function is greater for the lines decaying to the $^1D$ term than the $^3P$. Inspection of the components of the source function at 1500\,km shows that $\eta_\nu$ is only 50\% larger at 1114.46\,\AA\ than at 1256,50\,\AA\ and yet $\chi_\nu$ is a factor of 10 times larger at the shorter wavelengths by comparison, a clear indication of the greater opacity. It is just about possible to see that the contribution function of the line at 1256.50\,\AA\ is more intense than that of the line at 1114.46\,\AA. The $\tau=1$ layer for the longer wavelength line is around 1000\,km, whereas it is at 1500\,km for the shorter wavelength line. This leads to the greater intensity of the longer wavelength line seen in Fig.\;\ref{fig:rtlines}.

\begin{figure*}
\centering
\includegraphics[width=\textwidth]{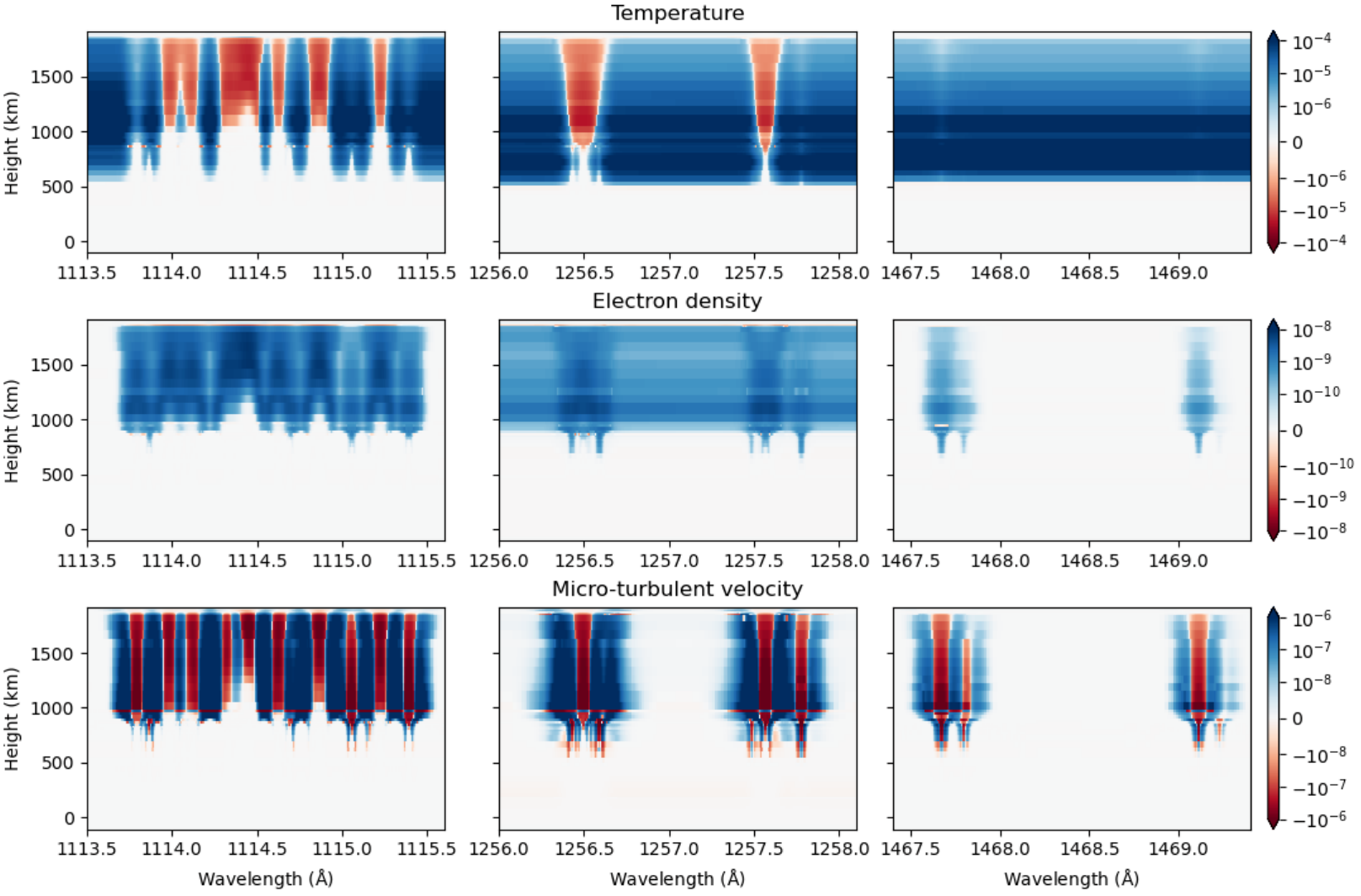}
\caption{Response functions for the lines emitted by the $2s^2\,2p\,10d,11s$ levels, following temperature, electron density and micro-turbulent velocity perturbations.}
\label{fig:n10_responses}
\end{figure*}

\subsection{Continuum formation}
\label{sec:cont}

Because the relative ion fractions of C$^0$ and C$^+$ are so important for the populations of the Rydberg levels and line intensities, we spend some time discussing the issues in modelling the ion fractions. The situation with C$^+$ can be understood through looking at the formation of the carbon continua, which is in emission and so is formed by radiative recombination from the C$^+$ ground state at these wavelengths. As found in \citet{dufresne2025lwatoms}, the three C$^0$ continuum edges are all lower in intensity than the SUMER observations when using the new atomic models. For the $^1D$ and $^1S$ continua the disagreement is slight, but a greater difference occurs with the $^3$P continuum. The continuum intensity from the old atomic models was higher than observations and yet the change in the C ion fractions between the new and old atomic models was reasonably small, indicating that the continuum is very sensitive to the ion fractions. 

\citet{dufresne2025lwatoms} discussed that the reason for the discrepancy in the continua could be caused by several factors. The decrease in the C$^+$ populations results, in part, from adding dielectronic recombination (DR) to the new atomic model for C. One factor that could explain the lower continua is that the DR rates from \citet{altun2004} are too high at these temperatures, decreasing the C$^+$ population. This scenario would, in turn, affect the intensities of the Rydberg lines because their populations are coupled to those of C$^+$. We explored this in the present work by using the C$^+$ DR rates from \citet{nussbaumer1983} and Si$^+$ rates from \citet{nussbaumer1986}. Both of those works paid careful attention to the rates at low temperatures. However, there was little change in the continua using these rates.

DR suppression should not be relevant at these temperatures even though the \citet{summers1972,summers1974} rates, from which we take the suppression, do not go below 10\,000\,K. \citet{dufresne2021picrm} show that including the new recombination rates from S$^+$ in the \citet{storey1995} collisional-radiative model makes it clear there is no DR suppression in the low-temperature component of the DR rates \citep{storey1981}. (This component occurs below 20\,000\,K for S$^+$.) They discuss on theoretical grounds that suppression should not occur for the low temperature component of DR. DR suppression is also discussed further in \citet{dufresne2025lwatoms} with regards to the effect the widely-used \citet{nikolic2013,nikolic2018} suppression factors would have on these elements at low temperatures.

Another factor causing the lower continua could be the model atmosphere stratification. The model atmosphere was semi-empirically adjusted to fit observations. If \citet{fontenla2014} had used atomic models of the type used by \citet{dufresne2025lwatoms} it may have produced different stratification in the atmosphere. Similarly, if, for example, \citet{fontenla2014} used other spectral features to adjust the atmosphere at the C$^0$ continua formation heights, it may have produced conditions which do not correctly reflect those that produce the C continua. We tested this further by using a different atmosphere, the enhanced network from \citet{fontenla2015}. This causes some features, such as the H and Si continua, to be brighter than observed, but it does show that with different C ion fractions the C continua and Rydberg line intensities are in very good agreement with observations. The synthetic spectra from this model are shown in Fig.\;\ref{fig:enhnetwork} and discussed in Appendix\;\ref{sec:enhnetwork} in more detail.

\subsection{Line response functions to atmospheric perturbations}
\label{sec:response_fns}

We show in Fig.\;\ref{fig:n10_responses} the response of the lines to perturbations in temperature, density and micro-turbulent velocities, the methods for which are described in Sect.\;\ref{sec:methods}. Again, for clarity, we highlight the changes for lines resulting from decays from the $2s^2\,2p\,10d,11s$ levels to all lower levels.  Increases in temperature usually increase the opacity of lines, and it is seen here for the lines around 1115\,\AA\ and 1257\,\AA\ that there are strong decreases in intensities in the cores of the stronger lines. It is noted that negative responses to temperature perturbations do not occur for the weaker lines in any of the series. Around all of the lines can be seen the response to the temperature fluctuations of the continuum, which shows strong enhancements. The responses of the lines near 1468\,\AA\ are too weak to show any clear change.

With regards to perturbations in density, the response is simpler because the level populations are linearly dependent on density. Thus, we only see positive responses to changes in density. The greatest changes occur where the contribution function peaks, close to the $\tau=1$ layer. The clearest response that the lines at around 1468\,\AA\ exhibit are with micro-turbulent velocity perturbations. The response of all three series to changes in micro-turbulent velocity is a decrease in the line core intensity and an enhancement in the line wings, as expected. Since the velocity perturbations are linear, a largely constant response is seen throughout the layers of the atmosphere above 1000\,km. This contrasts with the response to temperature and density fluctuations because fractional perturbations are used in those cases. It is clear in all three cases that the line responses are much smaller below 900\,km, that is, in the temperature minimum region. Although slight, it can just about be seen that the responses to velocity reverse below the temperature minimum region. There is no response from the continuum to velocity perturbations at any wavelength.

\section{Conclusions}
\label{sec:concl}

This study has shown in-depth how and where the neutral carbon Rydberg lines form in the solar chromosphere. They begin forming in the temperature minimum region and continue being emitted in the upper chromosphere. There is obvious opacity in the lines that form by decays to the $^3P$ ground state. It means they form higher up in the chromosphere than the lines that decay to the $^1D,\,^1S$ metastable terms. In addition, within each series the weaker lines start forming lower down than stronger lines; such variations are useful for diagnostic purposes if an instrument observes only a small spectral region. In this work we have only focussed on the lines emitted by levels which are in SB equilibrium with the ground of C$^+$, that is, for levels with $n\geq 10$ in the quiet Sun. They are much more convenient to model compared to lines from levels which are not in SB equilibrium. In the higher densities of active regions and flares it is possible that Rydberg levels with $n\geq 6$ may be in SB equilibrium, according to the models of \citet{storey1995}. Despite this, any Rydberg lines could provide useful diagnostics no matter the conditions in which they form, even if they are in non-LTE.

The present case has provided a useful test bed to assess the diagnostic potential of Rydberg lines. There are up to 200 lines emitted by neutral carbon from the three series investigated here which contribute to observable emission in the solar atmosphere. There are six atoms which emit such lines in the wavelength region 850-1600\,\AA. Since each atom is likely to form at different heights in the chromosphere the lines should probe a range of heights. In that case, these type of lines should provide a wealth of diagnostic tools for both theoreticians and observers.

One such use has been shown here through calculating response functions. The lines are clearly responsive to perturbations in atmospheric parameters like temperature, density and micro-turbulent velocities. They are potentially invaluable to those carrying out inversions of observational data, not least because their energy levels are in SB equilibrium. Lines which are emitted even in a small wavelength range form over a range of heights, providing good constraints for such reconstructions of the conditions which emitted them.

One of the main objectives of EUVST is to provide seamless observations from the chromosphere to the corona, to help identify the source region of the solar wind. In documents relating to the mission \citep{shimizu2021,shimizu2024}, it is stated that the instrument will observe plasma starting from 10\,000--20\,000\,K. Since it will observe carbon lines at wavelengths longer than 1115\,\AA\ the present study confirms that the beginning of its observing range will be at least as low as 6\,000\,K. Another feature of EUVST is its high spectral ($R=13500$, about 0.08\,\AA\ at 1100\,\AA) and spatial (0.4") resolution. Because the Rydberg lines form following recombination from C$^+$ it may be possible to test whether the lines have characteristics of the singly-charged ion or the atom. The wide spectral coverage of EUVST means that it could altogether make use of possibly hundreds of Rydberg lines for diagnostic purposes.

As the Rydberg levels involved in the formation of the spectral lines are in SB equilibrium, we believe that it is possible to construct an efficient inversion code suited to transitions of this type without solving the full non-LTE problem for all the lines of the series and background.
In the following we sketch the concept of this method.
Whilst a non-LTE solution is needed to determine the ion fractions for a given atmosphere, this can use a relatively simple atomic model with about 25 levels, and these fractions can likely be updated only every few atmospheric iterations.
Using these ion fractions, the response functions can be analytically computed in the style of \citet{ruizcobo_sir:1992}.
We stress that a careful treatment of background opacities will be necessary to handle lines in these wavelength regions, and efficient approximations following the work of \citet{dufresne2025lwatoms} will likely need to be developed.
Additionally, the series around 1240\,\AA{} will be affected by the wing of H Lyman-$\alpha$ (including non-equilibrium ionisation effects), thus requiring additional consideration. 
However, these lines are not within the wavelength ranges observed by EUVST, and so are not essential for its low temperature diagnostic potential.

\section*{Acknowledgements}

GDZ and RPD acknowledge support from STFC (UK) via the consolidated grants to the atomic astrophysics group at DAMTP, University of Cambridge (ST/T000481/1).
GDZ also acknowledges
support from STFC via the large grant award to the SAMS project (UKRI1165).

CMJO is grateful to the Royal Astronomical Society's Norman Lockyer fellowship, and the University of Glasgow's Lord Kelvin/Adam Smith Leadership fellowship for supporting this work.

Most of the atomic rates used in the present study and in \textsc{CHIANTI}
were produced by the UK APAP network
funded by STFC via several grants to the University of Strathclyde, led by the late
Nigel R. Badnell\footnote{https://doi.org/10.3390/atoms13060055}.

	\textsc{CHIANTI} is a collaborative project involving George Mason University, the University of Michigan, the NASA Goddard Space Flight Center (USA) and the University of Cambridge (UK). 

We thank the anonymous referee for useful comments on the manuscript.

\section*{Data Availability}

The level energies and radiative decay rates for the Rydberg states are taken from \citet{storey2023c1}. The atomic models for C, Si and S are available are taken from \citet{dufresne2025lwatoms}, while the data for the other atomic models are available in \citet{osborne2021lw}. The model atmospheres are from \citet{fontenla2014,fontenla2015}.



\bibliographystyle{mnras}
\bibliography{lw_rydberg} 




\appendix

\section{Testing the enhanced network atmosphere}
\label{sec:enhnetwork}

\begin{figure*}
\centering
\includegraphics[width=\textwidth]{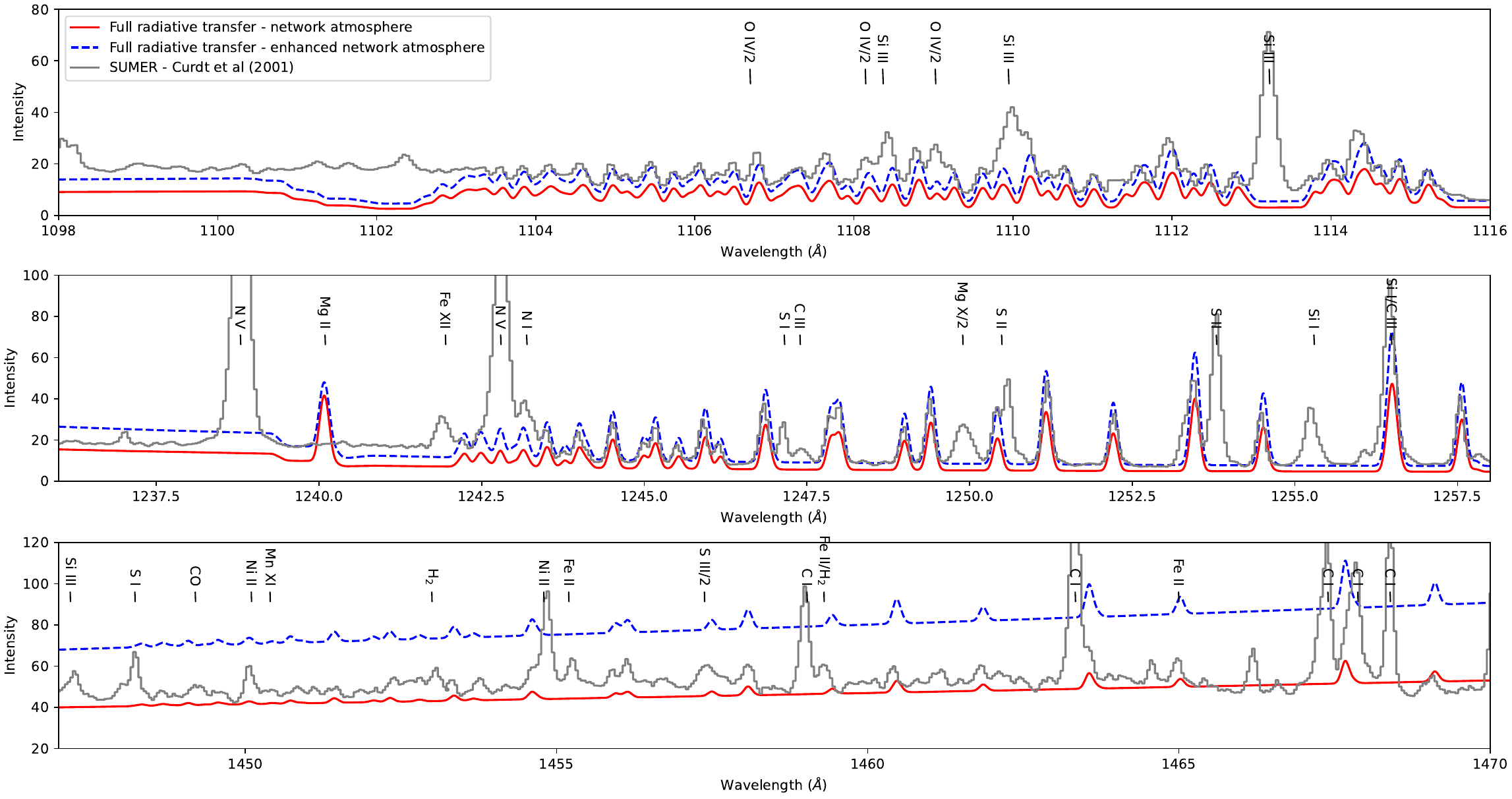}
\caption{Comparison of \ion{C}{i} Rydberg line intensities when using the enhanced network model atmosphere with the results from the network model and observations. The other main lines contributing to the spectrum that have been identified and are not being modelled here are indicated by black labels. The `/2' in the labels denotes a second order line in SUMER. All intensities are in erg\,cm$^{-2}$\,s$^{-1}$\,sr$^{-1}$\,\AA$^{-1}$.}
\label{fig:enhnetwork}
\end{figure*}

We continue here the discussion from Sect.\;\ref{sec:cont} about improving the results for the C continuum, the C$^+$ populations and, by consequence, the Rydberg lines. \citet{dufresne2025lwatoms} showed that the old C atomic model made available in \lw\ over-predicted the C$^+$ populations and continuum because DR was not included and the RR rates were too low. When the Rydberg lines are added to the ion fractions from the old atomic model their intensities are too strong by about a factor of two compared to observations, while the continuum is about 50\% higher. Figure\;\ref{fig:enhnetwork} compares the synthetic spectrum derived when using the \citet{fontenla2015} enhanced network model atmosphere in the present setup with the results from the network atmosphere. Clearly the continuum at 1100\,\AA, which is almost entirely emitted by C, is close to observations. It is also seen that the Rydberg line intensities are very close to observations in this wavelength range. The Rydberg series decaying to the $^1D$ states around 1245\,\AA\ are in better agreement as well. The only difference for this series is the continuum at shorter wavelengths, but this actually comes from the Lyman-$\alpha$ line, which is too broad and intense in the enhanced network atmosphere. Similarly, the continuum is too high at 1450\,\AA, but this is caused by an overly intense Si continuum. The few Rydberg lines in this series that are likely to be unblended in the observations are closer in intensity in the synthetic spectrum now, when subtracting the continua from theory and observations and only considering the line intensities themselves.

\begin{figure}
\centering
\includegraphics[width=\columnwidth]{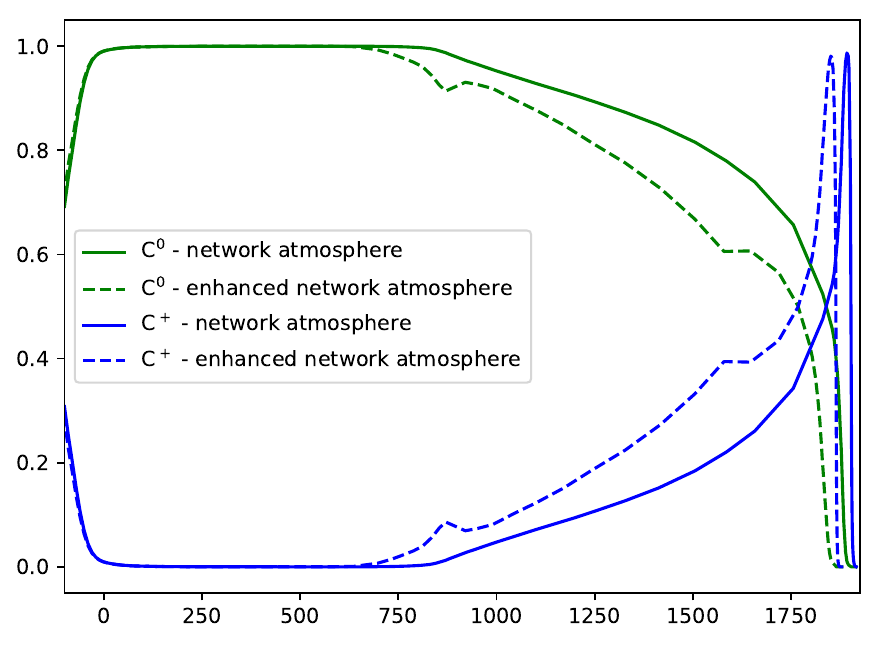}
\caption{Ion fractions for C$^0$ and C$^+$ with height for the network and enhanced network model atmospheres.}
\label{fig:ionfracs}
\end{figure}

It was suggested in Sect.\;\ref{sec:lines} that larger C$^0$ ion fractions could create too much opacity in the $^3P$ series at 1100\,\AA\ when using the network model. This could be causing the stronger lines at 1111.9\,\AA\ and 1114.4\,\AA\ to be proportionately weaker than other lines in the series. Figure\;\ref{fig:ionfracs} shows the ion fractions for C from the network and enhanced network atmospheres. It is clear that with the lower C$^0$ ion fractions in the enhanced model that the results for these particular lines are in better agreement with observations. None of this should be taken to mean that the enhanced network atmosphere is a better model; it is clear from other features that it does not reflect the present, quiet Sun observations. Its purpose here is only to highlight how sensitive the synthetic spectrum is to relatively small changes in the ion fractions and what effect they have on the lines. The Rydberg lines can be correctly modelled when using C$^+$ ion fractions which reproduce the observed continua.

\section{List of Rydberg lines emitted by the \texorpdfstring{\MakeLowercase{1s$^2$\,2s$^2$\,2p\,10d,11s\ }}\ states}

In Sects\;\ref{sec:contrfn}\;and\;\ref{sec:response_fns} we focus on a set of lines which are emitted from common upper levels with the configurations $1s^2\,2s^2\,2p\,10d,11s$. In Tab.\;\ref{tab:emiss} we list all the lines emitted by neutral carbon from these configurations. We also list the emissivity of the line to give an indication of the relative contribution by each line towards the observed intensities. The emissivity of a line is defined here as the energy emitted per unit time per C$^+$ ion for each line emitted at wavelength $\lambda_{ul}$ in a transition from upper level $u$ to lower level $l$ as

\begin{equation}
    \epsilon_{ul} \;=\; \frac{hc}{\lambda_{ul}} \; \frac{N_u}{N(C^+)} \, A_{ul} ~,
    \label{eqn:emiss}
\end{equation}

\noindent where $A_{ul}$ is the transition rate. For the present illustration the emissivity is calculated using the same conditions as \citet{storey2023c1}. They used a temperature of 7000\,K and density of 6$\times$10$^{10}$\,cm$^{-3}$, which are taken from the conditions at which \citet{lin2017} found the \ion{C}{i} 1355.8\,\AA\ line to form. The levels are designated by configuration, term and total angular momentum, $J$. Please see \citet{storey2023c1} for a description of the term notation and other details.

\begin{table*}
\caption{List of lines emitted by the $1s^2\,2s^2\,2p\,10d,11s$ states, which are the focus for the analysis later in the present work. $\lambda_{obs}$ and $\lambda_{calc}$ are the observed (if available) and theoretical wavelengths (in \AA) of the transition, respectively; $f_{\rm L}$ the absorption oscillator strength in the length form; $A_{ul}$ the transition rate (in s$^{-1}$); and, $\epsilon_{ul}$ is the emissivity (in ergs s$^{-1}$), as defined by Eq.\;\ref{eqn:emiss}, using a temperature of 7000\,K and density 6$\times$10$^{10}$\,cm$^{-3}$.}
\centering
\begin{tabular}{rrrrllllllr}
\hline
 $\lambda_{obs}$ & $\lambda_{calc}$ & $f_{\rm L}$ & $A_{ul}$  & Lower level & & & Upper level & & & $\epsilon_{ul}$ \\
 \hline

- & 1114.07 & 3.97E-04 & 2.13E+06 &               $2s^2 \, 2p^2$ &    $^3P$ & 1 & $ 2s^2 \, 2p (^2P^o_{3/2})\,11s $ &   2[3/2]$^o$ & 1 & 5.06E-16  \\
- & 1114.11 & 3.80E-05 & 1.22E+05 &               $2s^2 \, 2p^2$ &    $^3P$ & 1 & $ 2s^2 \, 2p (^2P^o_{3/2})\,11s $ &   2[3/2]$^o$ & 2 & 4.85E-17  \\
1114.13 & 1114.27 & 7.15E-04 & 2.31E+06 &               $2s^2 \, 2p^2$ &    $^3P$ & 1 & $ 2s^2 \, 2p (^2P^o_{3/2})\,10d $ &   2[5/2]$^o$ & 2 & 9.16E-16  \\
1114.33 & 1114.30 & 1.80E-07 & 1.61E+03 &               $2s^2 \, 2p^2$ &    $^3P$ & 2 & $ 2s^2 \, 2p (^2P^o_{3/2})\,10d $ &   2[3/2]$^o$ & 1 & 3.82E-19  \\
1114.33 & 1114.27 & 2.27E-04 & 2.03E+06 &               $2s^2 \, 2p^2$ &    $^3P$ & 2 & $ 2s^2 \, 2p (^2P^o_{3/2})\,10d $ &   2[1/2]$^o$ & 1 & 4.81E-16  \\
1114.33 & 1114.32 & 5.39E-04 & 2.89E+06 &               $2s^2 \, 2p^2$ &    $^3P$ & 2 & $ 2s^2 \, 2p (^2P^o_{3/2})\,10d $ &   2[3/2]$^o$ & 2 & 1.14E-15  \\
1114.38 & 1114.35 & 5.93E-04 & 2.28E+06 &               $2s^2 \, 2p^2$ &    $^3P$ & 2 & $ 2s^2 \, 2p (^2P^o_{3/2})\,10d $ &   2[7/2]$^o$ & 3 & 1.26E-15  \\
- & 1114.41 & 9.54E-05 & 8.54E+05 &               $2s^2 \, 2p^2$ &    $^3P$ & 2 & $ 2s^2 \, 2p (^2P^o_{3/2})\,11s $ &   2[3/2]$^o$ & 1 & 2.03E-16  \\
- & 1114.44 & 8.17E-04 & 4.39E+06 &               $2s^2 \, 2p^2$ &    $^3P$ & 2 & $ 2s^2 \, 2p (^2P^o_{3/2})\,11s $ &   2[3/2]$^o$ & 2 & 1.74E-15  \\
1114.46 & 1114.61 & 4.23E-05 & 2.27E+05 &               $2s^2 \, 2p^2$ &    $^3P$ & 2 & $ 2s^2 \, 2p (^2P^o_{3/2})\,10d $ &   2[5/2]$^o$ & 2 & 9.01E-17  \\
1114.46 & 1114.44 & 1.29E-03 & 4.93E+06 &               $2s^2 \, 2p^2$ &    $^3P$ & 2 & $ 2s^2 \, 2p (^2P^o_{3/2})\,10d $ &   2[5/2]$^o$ & 3 & 2.73E-15  \\
1114.62 & 1114.60 & 2.65E-03 & 4.74E+06 &               $2s^2 \, 2p^2$ &    $^3P$ & 0 & $ 2s^2 \, 2p (^2P^o_{1/2})\,10d $ &   2[3/2]$^o$ & 1 & 1.12E-15  \\
1114.69 & 1114.64 & 3.72E-04 & 6.66E+05 &               $2s^2 \, 2p^2$ &    $^3P$ & 0 & $ 2s^2 \, 2p (^2P^o_{1/2})\,11s $ &   2[1/2]$^o$ & 1 & 1.58E-16  \\
1114.83 & 1114.81 & 1.88E-04 & 1.01E+06 &               $2s^2 \, 2p^2$ &    $^3P$ & 1 & $ 2s^2 \, 2p (^2P^o_{1/2})\,10d $ &   2[3/2]$^o$ & 1 & 2.39E-16  \\
1114.87 & 1114.85 & 1.65E-03 & 5.32E+06 &               $2s^2 \, 2p^2$ &    $^3P$ & 1 & $ 2s^2 \, 2p (^2P^o_{1/2})\,10d $ &   2[3/2]$^o$ & 2 & 2.11E-15  \\
1114.88 & 1114.87 & 9.27E-05 & 1.49E+06 &               $2s^2 \, 2p^2$ &    $^3P$ & 1 & $ 2s^2 \, 2p (^2P^o_{1/2})\,11s $ &   2[1/2]$^o$ & 0 & 1.18E-16  \\
1114.90 & 1114.85 & 3.22E-05 & 1.73E+05 &               $2s^2 \, 2p^2$ &    $^3P$ & 1 & $ 2s^2 \, 2p (^2P^o_{1/2})\,11s $ &   2[1/2]$^o$ & 1 & 4.09E-17  \\
1115.06 & 1115.01 & 4.85E-05 & 1.56E+05 &               $2s^2 \, 2p^2$ &    $^3P$ & 1 & $ 2s^2 \, 2p (^2P^o_{1/2})\,10d $ &   2[5/2]$^o$ & 2 & 6.18E-17  \\
1115.16 & 1115.14 & 4.25E-06 & 3.80E+04 &               $2s^2 \, 2p^2$ &    $^3P$ & 2 & $ 2s^2 \, 2p (^2P^o_{1/2})\,10d $ &   2[3/2]$^o$ & 1 & 9.00E-18  \\
1115.20 & 1115.19 & 1.39E-05 & 7.43E+04 &               $2s^2 \, 2p^2$ &    $^3P$ & 2 & $ 2s^2 \, 2p (^2P^o_{1/2})\,10d $ &   2[3/2]$^o$ & 2 & 2.94E-17  \\
1115.23 & 1115.18 & 5.31E-05 & 4.75E+05 &               $2s^2 \, 2p^2$ &    $^3P$ & 2 & $ 2s^2 \, 2p (^2P^o_{1/2})\,11s $ &   2[1/2]$^o$ & 1 & 1.13E-16  \\
1115.23 & 1115.20 & 5.23E-04 & 2.00E+06 &               $2s^2 \, 2p^2$ &    $^3P$ & 2 & $ 2s^2 \, 2p (^2P^o_{1/2})\,10d $ &   2[5/2]$^o$ & 3 & 1.11E-15  \\
1115.39 & 1115.35 & 7.12E-05 & 3.82E+05 &               $2s^2 \, 2p^2$ &    $^3P$ & 2 & $ 2s^2 \, 2p (^2P^o_{1/2})\,10d $ &   2[5/2]$^o$ & 2 & 1.51E-16  \\

1256.42 & 1256.41 & 9.62E-07 & 4.07E+03 &               $2s^2 \, 2p^2$ &    $^1D$ & 2 & $ 2s^2 \, 2p (^2P^o_{3/2})\,10d $ &   2[3/2]$^o$ & 2 & 1.42E-18  \\
1256.43 & 1256.39 & 2.61E-04 & 1.84E+06 &               $2s^2 \, 2p^2$ &    $^1D$ & 2 & $ 2s^2 \, 2p (^2P^o_{3/2})\,10d $ &   2[3/2]$^o$ & 1 & 3.86E-16  \\
1256.43 & 1256.35 & 6.33E-08 & 4.46E+02 &               $2s^2 \, 2p^2$ &    $^1D$ & 2 & $ 2s^2 \, 2p (^2P^o_{3/2})\,10d $ &   2[1/2]$^o$ & 1 & 9.36E-20  \\
1256.50 & 1256.46 & 1.63E-03 & 4.93E+06 &               $2s^2 \, 2p^2$ &    $^1D$ & 2 & $ 2s^2 \, 2p (^2P^o_{3/2})\,10d $ &   2[7/2]$^o$ & 3 & 2.42E-15  \\
- & 1256.52 & 8.28E-05 & 5.83E+05 &               $2s^2 \, 2p^2$ &    $^1D$ & 2 & $ 2s^2 \, 2p (^2P^o_{3/2})\,11s $ &   2[3/2]$^o$ & 1 & 1.23E-16  \\
- & 1256.57 & 2.08E-08 & 8.77E+01 &               $2s^2 \, 2p^2$ &    $^1D$ & 2 & $ 2s^2 \, 2p (^2P^o_{3/2})\,11s $ &   2[3/2]$^o$ & 2 & 3.08E-20  \\
1256.59 & 1256.78 & 2.65E-05 & 1.12E+05 &               $2s^2 \, 2p^2$ &    $^1D$ & 2 & $ 2s^2 \, 2p (^2P^o_{3/2})\,10d $ &   2[5/2]$^o$ & 2 & 3.94E-17  \\
1256.59 & 1256.56 & 1.15E-04 & 3.48E+05 &               $2s^2 \, 2p^2$ &    $^1D$ & 2 & $ 2s^2 \, 2p (^2P^o_{3/2})\,10d $ &   2[5/2]$^o$ & 3 & 1.71E-16  \\
1257.49 & 1257.46 & 7.44E-06 & 5.23E+04 &               $2s^2 \, 2p^2$ &    $^1D$ & 2 & $ 2s^2 \, 2p (^2P^o_{1/2})\,10d $ &   2[3/2]$^o$ & 1 & 1.10E-17  \\
1257.54 & 1257.52 & 5.00E-06 & 2.11E+04 &               $2s^2 \, 2p^2$ &    $^1D$ & 2 & $ 2s^2 \, 2p (^2P^o_{1/2})\,10d $ &   2[3/2]$^o$ & 2 & 7.40E-18  \\
1257.57 & 1257.51 & 1.24E-04 & 8.72E+05 &               $2s^2 \, 2p^2$ &    $^1D$ & 2 & $ 2s^2 \, 2p (^2P^o_{1/2})\,11s $ &   2[1/2]$^o$ & 1 & 1.83E-16  \\
1257.57 & 1257.54 & 6.80E-04 & 2.05E+06 &               $2s^2 \, 2p^2$ &    $^1D$ & 2 & $ 2s^2 \, 2p (^2P^o_{1/2})\,10d $ &   2[5/2]$^o$ & 3 & 1.01E-15  \\
1257.78 & 1257.72 & 2.49E-05 & 1.05E+05 &               $2s^2 \, 2p^2$ &    $^1D$ & 2 & $ 2s^2 \, 2p (^2P^o_{1/2})\,10d $ &   2[5/2]$^o$ & 2 & 3.70E-17  \\

1467.67 & 1467.61 & 1.74E-03 & 1.79E+06 &               $2s^2 \, 2p^2$ &    $^1S$ & 0 & $ 2s^2 \, 2p (^2P^o_{3/2})\,10d $ &   2[3/2]$^o$ & 1 & 3.23E-16  \\
1467.67 & 1467.57 & 4.04E-04 & 4.17E+05 &               $2s^2 \, 2p^2$ &    $^1S$ & 0 & $ 2s^2 \, 2p (^2P^o_{3/2})\,10d $ &   2[1/2]$^o$ & 1 & 7.49E-17  \\
- & 1467.80 & 3.19E-04 & 3.29E+05 &               $2s^2 \, 2p^2$ &    $^1S$ & 0 & $ 2s^2 \, 2p (^2P^o_{3/2})\,11s $ &   2[3/2]$^o$ & 1 & 5.93E-17  \\
1469.11 & 1469.07 & 9.68E-04 & 9.97E+05 &               $2s^2 \, 2p^2$ &    $^1S$ & 0 & $ 2s^2 \, 2p (^2P^o_{1/2})\,10d $ &   2[3/2]$^o$ & 1 & 1.79E-16  \\
1469.23 & 1469.14 & 4.29E-05 & 4.42E+04 &               $2s^2 \, 2p^2$ &    $^1S$ & 0 & $ 2s^2 \, 2p (^2P^o_{1/2})\,11s $ &   2[1/2]$^o$ & 1 & 7.96E-18  \\

\hline
\end{tabular}
\label{tab:emiss}
\end{table*}


\bsp	
\label{lastpage}
\end{document}